%% akk Alex, 1959 - 2022: revisiting this may 2026 for arXiv
%% and yet other channels 

%% topoffile 

% From koning@few.eur.nl Thu Oct 28 09:39:18 1999
% Email addresses: nils@math.uio.no, koning@few.eur.nl
% check out http://www.gbhap.com/journals/718/718-nfc.htm

%% i need:
% nils_alex_1.ps
% nils_alex_2c.ps 
% nils_alex_3.ps 

% should follow this style: 
%  \smallskip
% \vskip-0.5cm 
% \centerline{\includegraphics[height=4.4in,width=3.3in,angle=270]
% {something_fig11.ps} }

% \footline={{\smallrm\quotation}
%        \hfil{\rm\the\pageno}\hfil{\smallrm\version, as of \today}}

\magnification\magstep1
\baselineskip14pt
\vsize=23.5truecm
\hsize15.25truecm
\hoffset0.60truecm

\input miniltx
\input graphicx

\font\csc=cmcsc10
\font\bigbf=cmbx12
\font\bigbf=cmbx12
\font\smallrm=cmr8

\def\hatt{\widehat}
\def\tilda{\widetilde}
\def\half{\hbox{$1\over2$}}
\def\eps{\varepsilon}
\def\RR{\mathord{I\kern-.3em R}}
\def\Var{{\rm Var}}
\def\E{{\rm E}}
\def\d{{\rm d}}

\def\diag{{\rm diag}}

\def\midd{{\,|\,}}

\def\sumin{\sum_{i=1}^n}
\def\section{\bigskip}
\def\subsection{\medskip}
\def\startit{\medskip\noindent}
\def\arr{\rightarrow}
\def\tr{{\rm t}}

\def\fermat#1{\setbox0=\vtop{\hsize4.00pc
        \smallrm\raggedright\noindent\baselineskip9pt
        \rightskip=0.5pc plus 1.5pc #1}\leavevmode
        \vadjust{\dimen0=\dp0
        \kern-\ht0\hbox{\kern-4.00pc\box0}\kern-\dimen0}}
\def\textref#1#2{{[#1]}}
% uncomment the next line for logging references in text
\def\textref#1#2{{[#1]}\immediate\write17{textref: [#1] #2}}

\centerline{\bigbf Tests for Constancy of Model Parameters Over Time}

\bigskip
\centerline{\bigbf Nils Lid Hjort and Alex J.~Koning}

\smallskip
\centerline{\bf University of Oslo and Erasmus Universiteit Rotterdam}

%\smallskip
%\centerline{\sl -- but this is only \version, as of \today{} --}

\bigskip
{\smallskip\narrower\baselineskip12pt\noindent\rm
{\csc Abstract.}
Suppose that a sequence of data points follows a distribution
of a certain parametric form, but that one or more of
the underlying parameters may change over time.
This paper addresses various natural questions
in such a framework. We construct canonical monitoring
processes which under the hypothesis of no change
converge in distribution to independent Brownian bridges,
and use these to construct natural goodness-of-fit statistics.
Weighted versions of these are also studied,
and optimal weight functions are derived to give
maximum local power against alternatives of interest.
We also discuss how our results can be used to pinpoint
where and what type of changes have occurred, in the event
that initial screening tests indicate that such exist.
Our unified large-sample methodology is quite general
and applies to all regular parametric models,
including regression, Markov chain and time series situations.

\smallskip\noindent\sl
{\csc Key words:}
Brownian bridges;
change points;
constancy of parameters;
goodness of fit testing;
parameter discontinuities
\smallskip}

\section
\centerline{\bf 1. Introduction and summary}

\startit
Do the parameters of a statistical model stay constant,
or do they experience changes over time?
What are the best goodness-of-fit tests for the `no change'
hypothesis? What is necessary in order to claim that
changes have occurred? If there are level shifts or
other types of discontinuity, how can one spot them,
or best describe their nature, or pinpoint their locations?

This paper is concerned with these general questions,
and aims at devising generally applicable principles
and methods.
% Our methodology with its constructive recipes,
% and the results obtained about these, is general enough to work
% in most parametric models -- including i.i.d.~models,
% regression models, time series and Markov chain models.
The basic theory is developed first in Sections 2--4,
for the structurally and conceptually simplest general case,
namely that of independent data with no extra covariate information.
Here $Y_1,Y_2,\ldots$ are independent with densities
of common form  $f(y,\theta)$, but the parameter
$\theta$ is not necessarily constant as time goes by.
In Section 2 a certain $p$-dimensional monitoring process
is constructed, $p$ being the dimension of the $\theta$ parameter,
behaving in the large-sample limit as $p$ independent
Brownian bridges. This makes it easy to construct various
overall tests for the hypothesis of no change in the $\theta$s,
having in mind as interesting alternatives those where
the parameter changes over time. In Section 3 weighted
versions of these processes are constructed.
Section 4 provides results about local detection power,
against various discontinuity alternatives of interest,
and about optimal weight functions.
To some extent we also learn about how to detect where
changes have occurred, if indeed the screening tests
indicate that such are present.

There would often be situations where covariate information
is available for each $Y_i$, and where questions
related to parameter constancy or change would be important.
As an example, suppose $Y_i$ is Poisson with
mean parameter $\exp(a+b_1x_{i,1}+b_2x_{i,2})$,
reflecting dependence on factors $x_1$ and $x_2$.
Then perhaps $b_2$ changes over time, reflecting
say increased dependence on factor $x_2$.
The general regression framework is discussed in Section 5.
Section 6 provides illustrations of our methods.

The scope of our methodology is broader than models
with independence. Tests for parameter constancy,
and results about these, may be derived also in more complex
situations, like in Markov chain models,
where the transition probabilities may have changed over time,
and in time series regimes, where for example serial
dependence parameters may not have been constant over time.
This is explained in Section 7, along with other
remarks and pointers to problems for further work.

There are several areas of applied statistics where
questions and problems arise for which the methods of
this paper would be applicable. One quite general such area
is that of prediction. This is of central importance
in econometrics, for example; 
see e.g.~\textref{16}{Ploberger and Kr\"amer (1992)}.
Ongoing debates and controversies concerning climate
changes also involve prediction issues.
Predictions rely heavily on the assumption that the future
behaves in much the same way as in the past. Structural
breaks may destroy the reliability of predictions.
The investigation of the Dutch Ombudsman, described in Section 6,
and where our methods do discover a structural break
in underlying parameters, was in fact initiated because
of the poor prediction of capacity needed for
so-called TBS-treatments.

Another general such area is statistical process control.
Statistical process control methods aim at detecting
non-constancy of parameters in the context of industrial statistics.
Our methods are of relevance for analysing historical data
(corresponding to what is sometimes referred to as
`stage I statistical process control' problems),
see \textref{22}{Sullivan and Woodall (1996)} and
\textref{13}{Koning and Does (1997)};
and also to some extent for monitoring real-time data.

And a third general area of application would be that
of stochastic simulation via processes that supposedly
converge towards equilibrium distributions.
Some simulation systems need a `warming-up'
period to reach stationarity. Some special cases
of tests presented in this paper are in fact
already used to investigate whether the system
has warmed up sufficiently; see
e.g.~\textref{20}{Schruben (1982)},
\textref{21}{Schruben (1983)}
and \textref{18}{Ripley (1987)}, Ch.~6.

It is worth remarking that the clear majority of articles
dealing with goodness of fit problems for parametric models
is concerned with a more `static' problem formulation;
one believes that a sample comes from a definite distribution
and tests whether this distribution is of a specified type.
The present formulation is `dynamic' and focusses specifically
on discontinuities over time. This helps explain why our large-sample
theory leads to results that are both more unified and
more simple than those obtained in the `static' framework.
Thus an infamous comment of \textref{17}{Pollard (1984)}, p.~118,
stating that ``The interest aroused when \textref{4}{Durbin (1973)} applied
weak convergence methods to get limit distributions for statistics
analogous to those of Kolmogorov and Smirnov, but with estimated
parameters, died down when the intractable limit processes
asserted themselves,'' does not concern us.

\section
\centerline{\bf 2. Canonical monitoring processes}

\startit
The framework for this section involves a sequence of independent
observations $Y_i$, coming all from the same parametric family
$f(y,\theta)$, where however the underlying parameters $\theta_i$,
all belonging to some open $p$-dimensional parameter region,
may have changed over time. After having observed $Y_1,\ldots,Y_n$,
we take particular interest in the hypothesis
$$H_0\colon \theta_1=\cdots=\theta_n, \eqno(2.1)$$
which is to be tested against `discontinuity alternatives'.
We assume standard regularity conditions hold for the
$f(y,\theta)$ family, sufficient to make the traditional
maximum likelihood apparatus work.

\subsection
{\sl 2.1. Cumulative score processes.}
Let $u(y,\theta)$ and $i(y,\theta)$ be first and second derivatives
of $\log f(y,\theta)$ w.r.t.~$\theta$.
To learn about possible evidence against $H_0$,
start out considering cumulative sums of $u(Y_i,\theta_0)$, where
$\theta_0$ is the common parameter value under $H_0$.
These have mean zero and variance matrix
$J=-\E i(Y,\theta_0)$, the information matrix of the model.
By the Donsker theorem, see e.g.~\textref{2}{Billingsley (1968)},
combined with the Cram\'er--Wold device,
it is not difficult to derive the result
$$\psi_n(t,\theta_0)={1\over \sqrt{n}}\sum_{i\le [nt]}u(Y_i,\theta_0)
        \arr_d Z_0(t) \quad {\rm in\ }D_p[0,1], \eqno(2.2) $$
where $Z_0$ is zero-mean Gau\ss ian with covariance function
$\min(t_1,t_2)\,J$. The convergence takes place w.r.t.~the
Skorohod topology in the space $D_p[0,1]$ of right-continuous
functions $x\colon[0,1]\arr{\RR}^p$ with left-hand limits.
Note that $Z_0$ is a linear transformation
of $p$ independent Brownian motions.

Our main concern will be with the case of unknown parameters
in the model. But it is worth pointing out that in
the fully specified case, where $H_0$ states that all
$\theta_i$s are equal to a specified $\theta_0$,
the component processes of $J^{-1/2}\psi_n(t,\theta_0)$
tend to $p$ independent Brownian motions under $H_0$.
This makes it particularly easy to construct and
analyse test statistics.

\subsection
{\sl 2.2. Estimated cumulative score processes.}
When $\theta_0$ is unknown, let $\hatt\theta$ be the
maximum likelihood estimator,
and consider the estimated cumulative score process:
$$\psi_n(t,\hatt\theta)
        ={1\over \sqrt{n}}\sum_{i\le [nt]}u(Y_i,\hatt\theta)
  \quad {\rm for\ }0\le t\le 1. $$
Notice that this process both starts and ends at zero.
We now use Taylor expansion in conjunction with well-known
results about the sampling behaviour of $\hatt\theta$,
e.g.~$\sqrt{n}(\hatt\theta-\theta_0)\doteq J^{-1}\psi_n(1,\theta_0)$,
where $A_n\doteq B_n$ means that $A_n-B_n$ tends to zero in
probability. With $u(Y_i,\hatt\theta)
\doteq u(Y_i,\theta_0)+i(Y_i,\theta_0)(\hatt\theta-\theta_0)$
this leads to
$$\eqalign{\psi_n(t,\hatt\theta)
&\doteq \psi_n(t,\theta_0)+{1\over n}\sum_{i\le [nt]}i(Y_i,\theta_0)
        \,\sqrt{n}(\hatt\theta-\theta_0) \cr
&\doteq \psi_n(t,\theta_0)-t J_{[nt]} J^{-1}\psi_n(1,\theta_0)
 \arr_d Z(t)=Z_0(t)-tZ_0(1), \cr}$$
where $J_n=-n^{-1}\sumin i(Y_i,\theta_0)$ is consistent for $J$.
The limit process $Z$ is a $p$-dimensional process
with covariance function $t_1(1-t_2)\,J$ for $t_1\le t_2$,
in other words a linear transformation
of $p$ independent Brownian bridges, see also
\textref{9}{Horv\'ath and Parzen (1994)}.

These results lead naturally to the construction of
{\sl the canonical monitoring process},
$$M_n(t)=\hatt J^{-1/2}\psi_n(t,\hatt\theta)
   =\hatt J^{-1/2}{1\over \sqrt{n}}\sum_{i\le [nt]}u(Y_i,\hatt\theta)
        \quad {\rm for\ }0\le t\le 1, \eqno(2.3)$$
where $\hatt J$ is any reasonable estimate of $J$
(in this connection, see also Remark 7.1).
The immediate and quite powerful result is then that
$$M_n\arr_d M=J^{-1/2}Z=(W^0_1,\ldots,W^0_p)^\tr
  \quad {\rm under\ }H_0, \eqno(2.4)$$
a vector with $p$ independent Brownian bridges
as component processes.

The inverse square root matrix is calculated as usual,
via eigen-analysis; there is an orthonormal
$P$ and a diagonal $D$ such that $P\hatt J P^\tr=D$,
containing eigenvalues of $\hatt J$ in decreasing order,
and one puts $\hatt J^{-1/2}=P^\tr D^{-1/2}P$.
Another option is the so-called LU-root.

\subsection
{\sl 2.3. Omnibus tests for constancy of parameters over time.}
Result (2.4) can easily be utilised for testing purposes,
as we demonstrate below. It is also quite useful
to monitor the component processes
$M_{n,j}(t)$ graphically, particularly in cases where
constancy has been rejected; such plots would help
in trying to pinpoint in which way or ways $H_0$ does not hold.
See Section 4.3 and the examples of Section~6.

\smallskip
{\csc Test 1:}
Classes of chi squared type tests can be developed
as follows. Divide $[0,1]$ into $m$ windows $I_1,\ldots,I_m$.
For component $j$, consider increments
$$\Delta M_{n,j}(I_k)={1\over \sqrt{n}}
        \sum_{i/n\in I_k}(\hatt J^{-1/2})_{(j)}\, u(Y_i,\hatt\theta). $$
These tend to $(\Delta W^0_j(I_1),\ldots,\Delta W^0_j(I_m))^\tr$.
Inverting the covariance matrix one finds that
$$A_{n,j}^2=\sum_{k=1}^m{\{\Delta M_{n,j}(I_k)\}^2\over |I_k|}
        \arr_d \chi^2_{m-1} \quad {\rm under\ }H_0, $$
where $|I_k|$ is the length of interval $I_k$.
These are component test statistics of separate use and interest.
They may also be combined to form one overall test, via
$$A_n^2=\sum_{j=1}^p A_{n,j}^2\arr_d \chi^2_{p(m-1)}
        \quad {\rm under\ }H_0. $$

\smallskip
{\csc Test 2:}
A $p$-dimensional Kolmogorov--Smirnov type test would be
$U_n=\max_{0\le t\le1}\|M_n(t)\|^2$, which can be written
$$\max_{0\le t\le1}\psi_n(t,\hatt\theta)^\tr\hatt J^{-1}
        \psi_n(t,\hatt\theta)
  ={1\over n}\max_{1\le j\le n-1}\Bigl(\sum_{i\le j}
   u(Y_i,\hatt\theta)\Bigr)^\tr
        \hatt J^{-1}\Bigl(\sum_{i\le j}u(Y_i,\hatt\theta)\Bigr). $$
Its limit distribution under $H_0$ is that of
$$\max_{0\le t\le 1}\|W^0(t)\|^2
        =\max_{0\le t\le 1}\Bigl\{\sum_{i=1}^p W^0_i(t)^2\Bigr\}^{1/2}. $$
Let us also point to a sum-of-Kolmogorov--Smirnov type tests option:
$$\eqalign{U_n'
&=\max_{0\le t\le 1}|M_{n,1}(t)|+\cdots+\max_{0\le t\le 1}|M_{n,p}(t)| \cr
&=\sum_{j=1}^p{1\over \sqrt{n}}\max_{1\le l\le n-1}\Big|\sum_{i\le l}
        (\hatt J^{-1/2})_{(j)}^\tr u(y_i,\hatt\theta)\Big|, \cr}$$
with limiting null distribution $\sum_{j=1}^p \max_{0\le t\le1}|W^0_j(t)|$.
The upper 0.05 quantile of the distribution of any one
these components is 1.358, for example; the distribution
of a sum of two or more such components can be found
via simulation.

We also mention the natural option of weighing by
inverse standard deviation. The point is that
$$T_{n,j}=\max_{\eps\le t\le 1-\eps}{|M_{n,j}(t)|\over \{t(1-t)\}^{1/2}}
  \arr_d \max_{\eps\le t\le 1-\eps}{|W^0_j(t)|\over \{t(1-t)\}^{1/2}}
  \quad {\rm under\ }H_0, $$
for each component $j$. This distribution can be simulated
or approximated, cf.~\textref{14}{Miller and Siegmund (1982)}.
Upper 0.10 and 0.05 quantiles are approximately 2.89 and 3.15,
for instance, for the case of $\eps=0.05$.
The $T_{n,j}$ would be calculated
as the maximum over all right- and left-hand limits
at points $t=k/n$ for which $\eps\le k/n\le 1-\eps$.

\smallskip
{\csc Test 3:}
As a final example of a general construction,
consider this $p$-dimen\-sional Cram\'er--von Mises type test:
$$C_n^2=\int_0^1\|M_n(t)\|^2\,\d t
  ={1\over n^2}\sum_{j=1}^{n-1}\Bigl(\sum_{i\le j}u(Y_i,\hatt\theta)\Bigr)^\tr
        \hatt J^{-1}\Bigl(\sum_{i\le j}u(Y_i,\hatt\theta)\Bigr). $$
Under $H_0$,
$$C_n^2\arr_d \sum_{j=1}^p\int_0^1 W^0_j(t)^2\,\d t
        =\sum_{k=1}^\infty{1\over \pi^2k^2}\chi^2_{k,p}. $$
Similarly, an Anderson--Darling type weighted version of this
can easily be put up.

\subsection
{\sl 2.4. Examples.}
The apparatus above, with monitoring processes and test
criteria built on these, can be routinely applied to
any regular parametric model.

\smallskip
{\csc Example 1:}
Assume the $Y_i$s come from a normal $(\mu,\sigma^2)$
distribution, where one at the outset could be interested
in monitoring both parameters for possible changes. Here
$u(y,\theta)=\sigma^{-1}(z,z^2-1)$, where $z=(y-\mu)/\sigma$,
and one quickly finds $J=\sigma^{-2}\diag(1,2)$.
This leads to
$$M_n(t)={1\over \sqrt{n}}\sum_{i\le [nt]}
  \pmatrix{Z_i \cr 2^{-1/2}(Z_i^2-1) \cr},
  \quad {\rm where\ } Z_i=(Y_i-\hatt\mu)/\hatt\sigma. $$
Tests for the constancy of $\mu$ or of $\sigma$ or both
can be constructed based on the two component processes,
as per the methods above. See also Remark 7.1.

The first component process coincides with the standardised time series
used in \textref{20}{Schruben (1982)}, \textref{21}{Schruben (1983)},
and is also a special case
of the least squares cumulative sum method in
\textref{16}{Ploberger and Kr\"amer (1992)}.

\smallskip
{\csc Example 2:}
Let the $Y_i$s come from a Gamma distribution with parameters
$(a,b)$, i.e.~with density $\{b^a/\Gamma(a)\}\,y^{a-1}\exp(-by)$.
Here one finds
$$M_n(t)=\pmatrix{\psi'(\hatt a) &-1/\hatt b \cr
                  -1/\hatt b & \hatt a/\hatt b^2 \cr}^{-1/2}
  {1\over \sqrt{n}}\sum_{i\le [nt]}
  \pmatrix{\log Y_i-\psi(\hatt a)+\log\hatt b \cr
           \hatt a/\hatt b-Y_i \cr}, $$
in terms of maximum likelihood estimators $(\hatt a,\hatt b)$.
The two component processes are again approximately
independent, and contribute combined information about
the mean level of $\log Y_i$ and the mean level of $Y_i$.
One may also construct monitoring processes focussing
on $a$ separately or $b$ separately, see the Remark below.

\smallskip
{\csc Example 3:}
Now take the $Y_i$s to be Poisson with parameters $\mu_i$.
The natural process to monitor these becomes
$M_n(t)=n^{-1/2}\sum_{i\le [nt]}(Y_i-\bar Y)/\bar Y^{1/2}$.
In the limit this is a Brownian bridge.

\smallskip
{\csc Example 4:}
Suppose a die is thrown many times. Assume that its
face probabilities $(p_1,\ldots,p_6)$ are unknown,
and imagine that they somehow may have changed over time.
To monitor this, let the data be registered
via $Y_i=(Y_{i,1},\ldots,Y_{i,6})$, with a 1 for the
face showing and 0 for the others. The recipe above gives
$$M_n(t)=\hatt J^{-1/2}{1\over \sqrt{n}}
  \sum_{i\le [nt]}\pmatrix{Y_{i,1}/\hatt p_1-Y_{i,6}/\hatt p_6 \cr
  \vdots \cr Y_{i,5}/\hatt p_5-Y_{i,6}/\hatt p_6 \cr}, $$
where $\hatt p_j=n^{-1}\sumin Y_{i,j}$ and where
the $5\times5$ matrix $J^{-1}$ has $p_j(1-p_j)$
along its diagonal and $-p_jp_k$ outside.
The five component processes are approximately independent
Brownian bridges, in the case that the probabilities
have been constant. Again various test statistics can
be written down as per Section 2.3.

Note that this example is relevant for the problem of
checking whether a probability density has changed
over time, via the monitoring of histograms.
More sophisticated methods for nonparametric monitoring
for changes of probability densities are 
given in~\textref{7}{Hjort and Koning (2000a)}.

\smallskip
{\csc Example 5:}
Let pairs $(X_i,Y_i)$ be independent and binormally
distributed, with parameters say $\mu_1,\sigma_1,\mu_2,\sigma_2,\rho$.
One may now construct a five-dimensional monitoring process
$M_n(t)$ whose limit distribution, under the hypothesis
of no change in the parameters, corresponds to five
independent Brownian bridges; we omit the algebraic details
here. One may then single out for example the fifth of these,
to look for possible changes in the $\rho$ parameter.

\smallskip
{\csc Remark:}
So far we have discussed monitoring processes in the
context of parametric models. One may also construct similar
methods to monitor statistical parameters more generally,
for example, nonparametrically checking the constancy
of the skewness parameter for an observed sequence.
Suppose $\alpha_i$ is such a parameter of interest,
connected to the distribution of $Y_i$, and that
the hypothesis $\alpha_1=\cdots=\alpha_n$ is to be
checked. Assume there is an estimator $\hatt\alpha_j$
for the common $\alpha$ value, depending on the first $j$
of $Y_i$ data, satisfying the standard requirement that
$\hatt\alpha_n-\alpha=n^{-1}\sumin I(Y_i)+R_n$,
where $I(y)$ is the influence function with variance say $\tau^2$,
and where $n^{1/2}R_n\arr_p0$. This ensures that
$A_n(t)=[nt]^{1/2}(\hatt\alpha_{[nt]}-\alpha)$
goes to a $N(0,\tau^2)$ for each positive $t$,
and more generally, under mild extra regularity,
that the process $A_n$ is asymptotically zero-mean Gau\ss ian
with covariance structure $(s/t)^{1/2}\tau^2$ for $s\le t$.
It follows that
$$B_n(t)=n^{-1/2}[nt](\hatt\alpha_{[nt]}-\hatt\alpha_n)
  =n^{-1/2}[nt]^{1/2}A_n(t)-n^{-1}[nt]A_n(1) $$
tends to $\tau W^0(t)$ in $D[0,1]$.
Hence $M_n(t)=B_n(t)/\hatt\tau$ is a Brownian bridge
in the large-sample limit, employing any reasonable
estimator $\hatt\tau$ of $\tau$.
This makes previous techniques apply for testing
the constancy hypothesis.

This apparatus may be used in the parametric models
above when there is a sub-parameter to be concentrated on,
like for example the shape parameter $\alpha$ in the Gamma model.
As another example, suppose pairs $(X_i,Y_i)$
are independent and that one is interested in monitoring their
correlation coefficients $\rho_i$. Let $\hatt\rho_j$
be the usual estimator based on the first $j$ pairs of data.
Then $n^{-1/2}[nt](\hatt\rho_{[nt]}-\hatt\rho_n)/\hatt\tau$
tends to a Brownian bridge under the hypothesis of no
change, for an appropriate scale estimate $\hatt\tau$.
Tests can now be constructed as above.
It should be pointed out that the convergence
to a Brownian bridge in this and similar examples,
though of course mathematically correct,
may be slower than for our (2.3) type processes,
particularly for smaller $t$.

\def\iovern{\hbox{${i\over n}$}}

\section
\centerline{\bf 3. Weighted monitoring processes}

\startit
The tests developed above are all of `omnibus type',
constructed without any particular attention to the types
of departures from $H_0$ that could be deemed more plausible
than others. When specific alternatives to constancy are envisaged
better tests can be constructed. This section works with
rich classes of goodness-of-fit processes, emerging by
integrating weight functions w.r.t.~the basic monitoring process.

Consider
$$V_n(t)=\int_0^t K_n(s)\circ \d M_n(s)
  ={1\over \sqrt{n}}\sum_{i\le [nt]}K_n(\iovern)
        \circ\hatt J^{-1/2}u(Y_i,\hatt\theta), $$
where $K_{n,1},\ldots,K_{n,p}$ are suitable weight functions,
and $a\circ b$ for two vectors indicates coordinate-wise multiplication.
In the interest of concise presentation we postpone
until the remark ending this section
discussing the exact regularity conditions needed for the
intended martingale and stochastic integration calculus to work;
these conditions at a minimum require the weight functions
to tend in probability to predictable functions $K_1,\ldots,K_p$.
Under such conditions one finds an appropriate generalisation of (2.4),
$$V_n\arr_d V, \quad
  {\rm with\ independent\ components\ }
  V_j=\int_0^t K_j(s)\,\d W^0_j(s), \eqno(3.1)$$
under $H_0$. These are again normal processes, and calculations yield
$${\rm cov}\{V_j(t_1),V_j(t_2)\}
        =\int_0^{t_1\wedge t_2}\!\!K_j^2\,\d s
                -\int_0^{t_1}\!K_j\,\d s\int_0^{t_2}\!K_j\,\d s. \eqno(3.2)$$

Many tests can be constructed using these $\int K_n\circ\,\d M_n$ processes,
along the lines of Section 2.3 for the special case of constant
weight functions. The difference is that the limit processes
become less tractable, but this is not a serious obstacle in view
of the practically simple option of simulating these when needed.
As an example, consider the supremum type test
which for component $j$ uses
$$U_{n,j}=\max_{0\le t\le 1}|V_{n,j}(t)|
  ={1\over \sqrt{n}}\max_{1\le l\le n}
        \Big|\sum_{i\le l}K_{n,j}(\iovern)(\hatt J^{-1/2})_{(j)}
        u(Y_i,\hatt\theta)\Bigr|, $$
which goes to $U_j=\max_{0\le t\le 1}|\int_0^t K_j\,\d W_j^0|$
under $H_0$. This distribution must then be approximated
by simulation. This remark applies also to sums over some or all
components, like $\sum_{j=1}^p U_{n,j}$, and to other
types of omnibus tests based on the $\int_0^t K_n\circ \d M_n$
processes.

A large class of tests which are more easily applied,
in that no simulation of limit distributions is called for,
is that of the chi squared tests.
Focus first on a single component, say $j$.
Divide again $[0,1]$ into $m$ cells $I_k$, and let
$$\Delta V_{n,j,k}=\int_{I_k}\d V_{n,j}(s)
        ={1\over \sqrt{n}}\sum_{i/n\in I_k}K_{j,n}(\iovern)
         (\hatt J^{-1/2})_{(j)}u(Y_i,\hatt\theta), $$
the increment over cell $I_k$. Then the vector of these,
under $H_0$, tends to the vector
$(\Delta V_{j,1},\ldots,\Delta V_{j,m})^\tr$, say,
which is zero-mean normal with covariance matrix of the form
$\Sigma_j=D_j-c_jc_j^\tr$. Here $D_j$ is diagonal with
$d_{j,k}=\int_{I_k} K_j^2\,\d s$
while vector $c_j$ has $c_{j,k}=\int_{I_k} K_j\,\d s$. And
$$\Sigma_j^{-1}=D_j^{-1}
  +D_j^{-1}c_jc_j^\tr D_j^{-1}/(1-c_j^\tr D_j^{-1}c_j), $$
giving the simple $\chi^2$ test
$$Q_{n,j}=\sum_{k=1}^m {(\Delta V_{n,j,k})^2\over \hatt d_{j,k}}
        +\Bigl(1-\sum_{k=1}^m{\hatt c_{j,k}^2\over \hatt d_{j,k}}\Bigr)^{-1}
         \Bigl(\sum_{k=1}^m{\hatt c_{j,k}\over \hatt d_{j,k}}
         \Delta V_{n,j,k}\Bigr)^2. \eqno(3.3)$$
Here $\hatt c_{j,k}$ and $\hatt d_{j,k}$ are natural
estimates of the respective quantities, typically using
$K_{n,j}$ instead of $K_j$. We have $Q_{n,j}\arr_d \chi^2_m$ under $H_0$,
unless $K=K_j$ is constant, which causes the final term to vanish and
a limiting $\chi^2_{m-1}$; see Test 1 of Section 2.4.

The above is valid for each of the components of $V_n$.
Summing over some or all components gives a grander test,
$$Q_n=\sum_{j=1}^p Q_{n,j}\arr_d \chi^2_{mp}
        \quad {\rm under\ }H_0. \eqno(3.4)$$
This holds since the individual transformed Brownian bridges are independent.

\smallskip
{\csc Remark:}
Various sets of regularity conditions can be put up
to ensure result (3.1). References include
\textref{5}{G\"annsler and H\"ausler (1979)},
\textref{19}{Rootz\'en (1980)}
and \textref{10}{Jacod and Shiryayev (1987)}.
The $K_{n,j}$ functions would either have to be
predictable (essentially, left-continuous processes
with values at time $s$ not depending on
outcomes of variables to be seen after time $s$),
or to be well enough approximated by predictable processes.
It would often suffice to have $K_{n,j}(s)$ of the form
$K_{n,j}(s,\hatt\alpha)$, where $\hatt\alpha$ is
$n^{1/2}$-consistent for a certain $\alpha$, and where
$K_{n,j}(s,\alpha)$ is predictable; see
\textref{6}{Hjort (1990)}, Section 2.1.
Next, $K_{n,j}(s)$ is required to converge in probability
to a predictable limit function $K_j(s)$, and we should have
$\int_0^t K_nK_n^\tr\,\d s\arr_p\int_0^t KK^\tr\,\d s$
for each $t$. Finally a Lindeberg type condition is
needed. We refer to \textref{19}{Rootz\'en (1980)} rather than spending
too many efforts discussing the details of his conditions
applied to our context. We note that Rootz\'en's methods and
conditions also apply to the regression framework of Section 5.

\section
\centerline{\bf 4. Local power and optimal weight functions}

\startit
The previous calculations have only been under the
constancy hypothesis. This section works out limiting
distribution results for various local alternatives,
and also derives the optimal form of the weight functions
when specific alternatives are being envisaged.
We also learn about the expected shapes of the monitoring
processes, under various alternatives. This is useful
when it comes to assessing the type of change that has occurred,
in cases where tests reveal that parameters have not been constant.

\subsection
{\sl 4.1. Limiting distributions for local alternatives.}
Consider alternatives in the vicinity of $H_0$, of the form
$\theta_i=\theta_0+\delta\circ h(\iovern)/n^{1/2}+O(1/n)$,
for departure functions $h=(h_1,\ldots,h_p)^\tr$ of suitable shapes
and degrees of departure $\delta=(\delta_1,\ldots,\delta_p)^\tr$.
Then $Y_i$ comes from
$$f(y,\theta_i)\doteq f(y,\theta_0)\,\{1
        +u(y,\theta_0)^\tr\,\delta\circ h(\iovern)/\sqrt{n}\}. \eqno(4.1)$$
Consider again the basic score process
$\psi_n(t,\theta_0)=n^{-1/2}\sum_{i\le [nt]}u(Y_i,\theta_0)$.
Presently it has mean function
$$\E\psi_n(t,\theta_0)
\doteq {1\over n}\sum_{i\le [nt]}\int f(y,\theta_0)
        u(y,\theta_0)u(y,\theta_0)^\tr\,\d y\,\delta\circ h(\iovern)
\arr J\int_0^t \delta\circ h(s)\,\d s. $$
The covariance function is different from what it is under $H_0$,
but only by an $O(1/n)$ effect. Further details ensure
$\psi_n(t,\theta_0)\arr_d J\int_0^t \delta\circ h(s)\,\d s+Z_0(t)$,
where $Z_0$ is as with $\delta=0$, that is, it is
zero-mean Gau\ss ian with covariance function $\min(t_1,t_2)\,J$.

Convergence of the estimated cumulative score process
can now be assessed outside the null hypothesis.
Some analysis, appropriately generalising arguments
used in Section 2, leads to
$$\eqalign{\psi_n(t,\hatt\theta)
&=\psi_n(t,\theta_0)-t\psi_n(1,\theta_0)+o_p(1) \cr
&\arr_d J\Bigl(\int_0^t\!\delta\circ h\,\d s
        -t\int_0^1\!\delta\circ h\,\d s\Bigr)+Z_0(t)-tZ_0(1). \cr}$$
It follows that
$$M_n(t)=\hatt J^{-1/2}\psi_n(t,\hatt\theta)
  \arr_d J^{1/2}\int_0^t\delta\circ (h-\bar h)\,\d s + W^0(t), \eqno(4.2)$$
under $\theta_i=\theta_0+\delta\circ h(\iovern)/\sqrt{n}$ circumstances.
Here $\bar h=\int_0^1 h(s)\,\d s$, and $W^0$ is a vector
of $p$ independent Brownian bridges $W^0_j$.
The above generalises result (2.4), which corresponds to
the case of $h$ being constant.

Without going too much into the technical details we
note that the methods and results of Rootz\'en and others,
as explained in the remark ending the previous section,
yield results for the weighted processes $\int_0^tK_{n,j}\,\d M_{n,j}(s)$
studied in Section 3, in the present local alternatives framework.
Thus for each $V_{n,j}$ we have, under (4.1) circumstances,
$$\int_0^t K_{n,j}(s)\,\d M_{n,j}(s)
  \arr_d \int_0^t K_j(s)(J^{1/2})_{(j)}\delta\circ(h(s)-\bar h)\,\d s
  +\int_0^t K_j(s)\,\d W^0_j(s) \eqno(4.3)$$
modulo regularity assumptions discussed before.
Result (4.3) generalises that of (3.1).
% $$\eqalign{V_{n,j}(t)
% &=\int_0^t K_{n,j}(s)\,\d M_{n,j}(s) \cr
% &\arr_d \int_0^t K_j(s)(J^{1/2})_{(j)}\delta\circ(h(s)-\bar h)\,\d s
%   +\int_0^t K_j(s)\,\d W^0_j(s) \cr} \eqno(4.3)$$
%$$V_n(t)=\int_0^t K_n\circ \d M_n(s)\arr_d
%        \int_0^t K(s)\circ \{J^{1/2}
%        \delta\circ(h(s)-\bar h)\}\,\d s+\int_0^t K\circ\d W^0, \eqno(4.3) $$

\subsection
{\sl 4.2. Weight functions and optimal local power.}
Using results above one may calculate approximate power
for various tests based on $M_n$ and $\int K_n\,\d M_n$,
against various alternatives of interest.
Given departure functions $h_1,\ldots,h_p$,
results (4.2)--(4.3) lead to expressions for limiting power,
depending on the degrees of departure $\delta_1,\ldots,\delta_p$.
This approach is rather complicated but nevertheless
quite useful when it comes to exploring the performance
of several of the supremum and integration based tests
portrayed in Section 2.3. It lends itself most easily
to the chi square type tests, however.

Consider the local power of the chi squared tests
constructed in Section 3. Focus on a single component $j$ first.
Now study the $\chi^2$ tests based on quadratic forms in
$\Delta V_{n,j,k}=\int_{I_k}K_{n,j}(s)\,\d M_{n,j}(s)$.
The vector of such increments tends to
$$\Bigl(\int_{I_1}K_jH_j\,\d s+\Delta V_{j,1},\ldots,
        \int_{I_m}K_jH_j\,\d s+\Delta V_{j,m}\Bigr)^\tr, $$
where $H_j(s)=(J^{1/2})_{(j)}\delta\circ (h(s)-\bar h)$.
For the test of (3.3) one therefore finds a noncentral chi squared limit,
$$Q_{n,j}\arr_d \chi^2_m(\lambda_j),
  \quad {\rm where\ }\lambda_j=a_j^\tr(D_j-c_jc_j^\tr)^{-1}a_j, \eqno(4.4)$$
% in the notation explained around (3.3),
with $a_{j,k}=\int_{I_k}K_jH_j\,\d s$.
The excentre parameter $\lambda_j$ can also be written
$$\sum_{k=1}^m {(\int_{I_k} K_jH_j\,\d s)^2\over \int_{I_k}K_j^2\,\d s}
+\Bigl(1-\sum_{k=1}^m{(\int_{I_k} K_j\,\d s)^2
        \over \int_{I_k} K_j^2\,\d s}\Bigr)^{-1}
        \Bigl(\sum_{k=1}^m{\int_{I_k} K_j\,\d s\over \int_{I_k} K_j^2\,\d s}
        \int_{I_k} K_jH_j\,\d s\Bigr)^2. $$
%$$Q_{n,j}\arr_d (a_{j,1}+\Delta V_{j,1},\ldots,a_{j,m}+\Delta V_{j,m})
%  (D_j-c_jc_j^\tr)^{-1}
%  (a_{j,1}+\Delta V_{j,1},\ldots,a_{j,m}+\Delta V_{j,m})^\tr $$
%$$Q_{n,j}=\sum_{k=1}^m {(\Delta V_{n,k})^2\over \hatt d_k}
%        +\Bigl(1-\sum_{k=1}^m{\hatt c_k^2\over \hatt d_k}\Bigr)^{-1}
%         \Bigl(\sum_{k=1}^m{\hatt c_k\over \hatt d_k}\Delta V_{n,k}\Bigr)^2
%         \arr_d \chi^2_m(\delta^2\lambda). \eqno(4.4)$$
The bigger $\lambda_j=\lambda_j(K_j)$, the greater power of the tests.
The optimal choice of $K_j$ can be proved to be
$$K_j(s)=H_j(s)=(J^{1/2})_{(j)}\{\delta\circ(h(s)-\bar h)\} \eqno(4.5)$$
(or proportional to this choice), see the appendix.
It attains the maximum possible value
$\lambda_j(H_j)=\max_{K_j} \lambda_j(K_j)=\int_0^1 H_j(s)^2\,\d s$,
and the corresponding optimal local power in (4.4).

The above is valid for each component of the vector
$V_n(t)=\int_0^t K_n\circ \d M_n$.
For combined $\chi^2$ tests the optimal weight function
is to make $K_n$ proportional to a consistent estimate of
$$K(s)=J^{1/2}\{\delta\circ (h(s)-\bar h)\}, $$
where $h_1,\ldots,h_p$ are the departure functions of interest.
This gives $\chi^2_{mp}$ tests of the form (3.4),
with limiting maximal power determined by the excentre parameter
for the $\chi^2_{mp}$ distribution, namely
$$\lambda=\sum_{j=1}^p \int_0^1 \bigl\{(J^{1/2})_{(j)}
        (\delta_jh_j(s)-\delta_j\bar h_j)\bigr\}^2\,\d s
  =\int_0^1 (\delta\circ (h-\bar h))^\tr\,J(\delta\circ(h-\bar h))\,\d s. $$
For comparison, the simpler test statistic using constant
$K_j$ functions, corresponding to $A_n^2$ of Section 2.3,
has a $\chi^2_{(m-1)p}(\mu)$ limit distribution with
excentre parameter $\sum_{j=1}^p\sum_{k=1}^m(\int_{I_k}H_j\,\d s)^2/|I_k|$.

\subsection
{\sl 4.3. Shape of $M_n$ and $V_n$ plots.}
Result (4.2) also provides useful information about
the expected shape of $M_n$ plots under different circumstances.
Take the $\delta_j$s to be equal, for simplicity.
We see then that the expected $M_{n,j}$ plot is proportional to
$(J^{1/2})_{(j)}$ times the vector of $\int_0^t(h-\bar h)\,\d s$.
If $h_j$ is a change point departure function,
say zero on $[0,a)$ and then equal to $b$ on $[a,1]$,
this integral is proportional to the triangular
function with value $-(1-a)t$ and $a(t-1)$ on respectively
$[0,a]$ and $[a,1]$. If on the other hand $h_j$ describes
a linear trend of change, as in $h_j(s)=cs$, then
the expected shape of the appropriate monitoring component
is $-\half ct(1-t)$, a symmetric parabola.
Illustrations 1 and 2 of Section 6 give examples of such behaviour.

It is also possible to estimate the position of break
points, for any of the parameters, in cases where initial
tests indicate that the (2.1) hypothesis does not hold.
Suppose for simplicity of discussion that there is only
one parameter to consider, and that this parameter
has a jump at an unknown position $a$. Then $M_n$ can
be represented as a triangular function plus noise,
as explained above. An estimate of $a$ emerges by
fitting the $M_n$ to a triangular shape and looking for
its top point.

If the alternative to $H_0$ of (2.1) is that of a linear
trend, say $\theta_i=\theta_0+(i/n)c$, then the optimal
weight function to use, by result (4.5), is $K(s)=s-\half$.
Thus $V_n(t)=\int_0^t(s-\half)\,\d M_n(s)$ is most
readily detecting the existence of such a trend.
The limiting null distribution of $\max_t|V_n(t)|$
is that of $\max_t|\int_0^t(s-\half)\,\d W^0(s)|$,
for example. Simulations showed that this distribution
has median about 0.32 and upper 0.05 quantile point about 0.64,
for example. In brief simulations for the situation
examined in Illustration 2 of Section 6 below,
the $\max_t|V_n(t)|$ test was indeed consistently better
than the $\max_t|M_n(t)|$ test. This also suggests additional
plots to be constructed for special tasks, like
plotting $V_n(t)/\tau(t)$ in the mentioned situation, where
$\tau(t)^2={1\over3}(t-\half)^3+{1\over 24}
   -{1\over4}\{(t-\half)^2-{1\over 4}\}^2$
is the limiting variance, as per (3.2).
\eject

\section
\centerline{\bf 5. Regression models}

\startit
There are many situations where observations $Y_i$ have
relevant covariate information $x_i$, and where interest
would focus on whether the precise form in which
the distribution of $Y_i$ depends on $x_i$
somehow could have changed over time.
Under suitable assumptions the previous no-covariate
methodology can be readily extended to the regression framework.

We take the $Y_i$s to be conditionally independent
given the sequence of $x_i$s, and will analyse sampling
behaviour in such a conditional framework, considering
$x_1,\ldots,x_n$ as given. We shall however assume that
the $x_i$s arrive as an exchangeable or ergodic sequence,
where averages stabilise in probability;
they could for example themselves
be i.i.d.~outcomes from some random mechanism.

Assume that $Y_i$ given $x_i$ comes from a density of
the form $f(y_i\midd x_i,\theta_i)$.
Let $u(y\midd x,\theta)$ and $i(y\midd x,\theta)$
be the first and second derivatives of $\log f(y\midd x,\theta)$
with respect to the parameters. Under model conditions
and the hypothesis $H_0$ that the parameters $\theta_i$
do not change, there is process convergence
$$\psi_n(t,\theta_0)={1\over \sqrt{n}}\sum_{i\le [nt]}
  u(Y_i\midd x_i,\theta_0)
  \arr_d Z_0(t) \quad {\rm in\ }D[0,1], \eqno(5.1)$$
under mild regularity conditions. Here $\theta_0$ is
the common true parameter value. The variance matrix of
$\psi_n(t,\theta)$ is $n^{-1}[nt]J_{[nt]}$, where
$$J_n=n^{-1}\sumin V(x_i,\theta_0)\arr_p J
  \quad {\rm as\ }n\arr\infty, \eqno(5.2)$$
writing $V(x_i,\theta_0)$ for $\Var\,u(Y_i\midd x_i,\theta_0)$
(assumed to be finite). That $J_n$ stabilises follows
from our ergodicity assumption about the $x_i$ sequence.
It follows from this, and the Lindeberg-extended Donsker theorem,
that the $Z_0$ limit has independent increments with
covariance structure $\min(t_1,t_2)J$.

Likewise other arguments of Section 2 can be utilised
and generalised to the present framework, showing that
if $\hatt\theta$ is the maximum likelihood estimator, then
$$\eqalign{\psi_n(t,\hatt\theta)
  &={1\over \sqrt{n}}\sum_{i\le [nt]} u(Y_i\midd x_i,\hatt\theta) \cr
  &\doteq \psi_n(t,\theta_0)-tJ_{[nt]}J^{-1}\psi_n(1,\theta_0)
   \arr_d Z(t)=Z_0(t)-tZ_0(1). \cr}$$
Again, this process is Gau\ss ian with zero mean and
covariance structure $t_1(1-t_2)J$ for $t_1\le t_2$.
And a canonical monitoring process emerges, of the form
$$M_n(t)=\hatt J^{-1/2}\psi_n(t,\hatt\theta)
  \quad {\rm for\ }0\le t\le 1. \eqno(5.3)$$
Here $\hatt J$ is any reasonable estimator of $J_n$,
and required to be consistent for the limiting matrix $J$
as $n$ grows; a natural choice is $n^{-1}\sumin V(x_i,\hatt\theta)$.
The limiting process $J^{-1/2}Z$ is that of $p$ independent Brownian
bridges, where $p$ is the number of parameters in $\theta$.

We note that the constructions and results
of weighted monitoring processes, discussed in Sections 3 and 4,
can be extended to the regression case, partly in view
of the methodology of \textref{19}{Rootz\'en (1980)} and
\textref{5}{G\"annsler and H\"ausler (1979)}.

\smallskip
{\csc Example 1:}
Let $Y_i$ given $x_i$ be normal $(x_i^\tr\beta,\sigma^2)$,
where $x_i$ is $p$-dimensional and $\beta_1,\ldots,\beta_p,\sigma$
are unknown parameters. Then
$$J_n=\sigma^{-2}\pmatrix{n^{-1}\sumin x_ix_i^\tr &0 \cr 0 & 2 \cr}, $$
and the monitoring vector process takes the form
$$M_n(t)={1\over \sqrt{n}}\sum_{i\le [nt]}
  \pmatrix{Z_ix_i\cr 2^{-1/2}(Z_i^2-1) \cr},
  \quad {\rm where\ }Z_i=(Y_i-x_i^\tr\hatt\beta)/\hatt\sigma. $$
This appropriately generalises the process of Example 1, Section 2.4.
The last component process of $M_n$ can be used to look for
changes in the $\sigma$ parameter, for example;
see the second illustration of Section 6.

\smallskip
{\csc Example 2:}
Let $Y_i$ be Poisson with mean parameter $\exp(x_i^\tr\beta)$,
and let $\hatt\beta$ be the maximum likelihood estimator.
The monitoring process takes the form
$$M_n(t)=\Bigl\{n^{-1}\sumin \exp(x_i^\tr\hatt\beta)\,x_ix_i^\tr\Bigr\}^{-1/2}
  {1\over \sqrt{n}}\sum_{i\le [nt]}\{Y_i-\exp(x_i^\tr\hatt\beta)\}x_i
  \quad {\rm for\ }0\le t\le 1. $$

\smallskip
{\csc Remark:}
There are situations where ergodicity of the $x_i$s
cannot be assumed, for instance in the case where
one of the covariates in question is the running time $i$
itself. Without ergodicity one is faced with additional
problems with no clear-cut general-purpose solution.
The variance function of the score process $\psi_n(t,\theta_0)$
may not factorise into a scalar function of $t$
and the information matrix $J$, which precludes a global standardisation
(that is, premultiplying by $J^{-1/2}$) of the score increments
$u(Y_i\midd x_i,\theta)$. Instead, a local standardisation
should be used. However, the obvious candidate
for such a standardisation, premultiplying
by $V(x_i,\theta_0)^{-1/2}$, is not applicable if
$V(x_i,\theta_0)$ is not of full rank.
% For instance, consider Example 1 with $p=2$ and $x_i=(1~i)^T$.
Further work is needed to solve such problems in
a satisfactory manner. Note however that if $\theta$ is scalar,
these problems do not emerge since the variance function
of the score process trivially factorises.

\section
\centerline{\bf 6. Illustrations and applications}

\startit
It is easy to illustrate the behaviour of our
canonical and weighted monitoring processes,
under the hypothesis of no change as well as under
various discontinuity alternatives, via simulations
from models of interest. We briefly provide some such
as Illustrations 1 and 2 below. In addition an application
is described using data from the Dutch Ombudsman,
looking for changes in Poisson model parameters.

\subsection
{\csc Illustration 1:}
Suppose data $Y_i$ come from a Gamma $(a_1,b_1)$
for $i=1,\ldots,100$ and from another Gamma $(a_2,b_2)$
for $i=101,\ldots,200$. We take here the mean levels
$a_1/b_1$ and $a_2/b_2$ to be the same, but scale
up the standard deviation $a_2^{1/2}/b_2$ to be
1.25 times that of $a_1^{1/2}/b_1$. It is not
easy to spot from just a plot of the 200 data points
that anything has happened to the underlying parameters.
However, the first component of the
monitoring process $M_n(t)$, see Example 2 of Section 2.4,
signals by its triangular shape and maximum size that something
has happened around $t=\half$, that is, around data point
100 of the 200. The upper 0.05 quantile of the distribution
of $\max_{0\le t\le 1}|W^0(t)|$ is 1.358, so $M_{n,1}$
clearly does not agree with the Brownian bridge behaviour
it should have had under the hypothesis of no change.

\smallskip
\centerline{\sl --- Figure 1 around here, see page 21 ---}

\subsection
{\csc Illustration 2:}
Consider a regression situation where data $Y_i$
are normal $(a+bx_i,\sigma^2)$, where the regression
coefficients do not change, but where $\sigma_i$
slowly increases with time. Specifically, we simulate
200 points around the regression line $1.11+2.22\,x$,
with $x$s being uniform on the unit interval,
and with $\sigma_i=1+0.5\,i/200$, increasing linearly from 1 to 1.5.
It is quite difficult to spot from scatterplots
or residuals that the standard deviation has been
increasing linearly over time.
But a look at the three monitoring plots quickly shows
that the third component reaches outside $\pm1.358$,
the 0.95-probability band, and that the two first components,
corresponding to the $a+bx$ part, stay nicely within.
This time the third monitoring process approximately forms
a parabola departure from zero, indicating as per the theory
of Section 4.3 that the non-constancy of the $\sigma$ parameter
might be in the form of a linear trend.
As explained there alternative plots
involving $V_{n,3}(t)=\int_0^t(s-\half)\,\d M_{n,3}(s)$,
not shown in our article, are even better at detecting
linear trends in the $\sigma$ parameter.

\smallskip
\centerline{\sl --- Figure 2 around here, see page 22 ---}

\subsection
{\csc An application:}
The first paragraph of article 78a of the Dutch Constitution reads
as follows: ``On request or on his own initiative the National
Ombudsman shall investigate the actions of administrative
authorities of the national government and of other administrative
authorities designated by or pursuant to Act of Parliament.''

In the Netherlands, convicted criminals may receive psychiatric treatment
in so-called TBS-institutions as part of their sentence.
% TBS officially means `terbeschikkingstelling met bevel
% tot verpleging van overheidswege', but I think
% `TerBeschikkingStelling' is sufficient. It means
% `to be put at the disposal of [of the authority, for psychiatric treatment].
(This Dutch acronym for `terbeschikkingstelling' indicates
in this case `to be put at the disposal of',
by the authority, for psychiatric treatment.)
The psychiatric treatment precedes the actual prison sentence.
Criminals on a waiting list for placement in a TBS-institution
are temporarily imprisoned under relatively poor conditions.
After receiving eleven complaints between December 1995 and February 1996,
the National Ombudsman decided to investigate on his own
initiative the TBS waiting lists, and especially
the waiting time involved. The investigation was reported
on in \textref{15}{National Ombudsman (1996)}. In Table I
% presented in Appendix II
the number of TBS-sentences and the number of ended TBS-treatments
are given for each month during the years 1984--1992.
Figure 3 displays the corresponding monitoring plots.

\smallskip
\centerline{\sl --- Figure 3 about here, see page 23 ---}

\smallskip
The monitoring plot of the expected number of ended treatments
exceeds the value 1.358, indicating that the hypothesis of
constancy of this parameter should be rejected at the 5\%
significance level. Moreover, the plot resembles a triangular
shape reaching its maximum deviation from the time axis in
March 1990; as explained in Section 4.3, this is indicative of
a change point. The plot suggests that around March 1990,
there was a sudden decrease in the expected number of ended treatments.
A possible explanation could be the increased complexity
of the psychiatric problems of the clients within the TBS-system.
Due to several policy changes in Dutch psychiatric care in the
late eighties, it became easier for unwilling psychiatric patients
to avoid admittance to psychiatric institutions. For these
patients (among them extreme psychotic patients) the TBS-system
started to act as a dust-bin.
%
% Alex encountered the description `zorgwekkende zorgmijders'
% somewhere, unfortunately I do not know how to translate!
% The direct translation is `troublesome care avoiders'.

\section
\centerline{\bf 7. Supplementing remarks}

\startit
This section lists various comments pertaining to the
use and further study of our methods.

\subsection
{\sl 7.1. Model-robust variance matrix estimation.}
In the framework of Section 2, suppose that under $H_0$
there is indeed a common density $f(y)$ for the $Y_i$s,
but that this unchanged density is not a member of the
parametric class $f(y,\theta)$. Still the maximum
likelihood estimator is meaningful, taking aim at the
least false parameter value $\theta_0$ which minimises
the Kullback--Leibler distance from the true to the
parametrised density, and there is convergence in
distribution $\sqrt{n}(\hatt\theta-\theta_0)$
towards a normal $(0,J^{-1}KJ^{-1})$. Here
$J=-\E_fi(Y_i,\theta_0)$ and $K=\Var_f\,u(Y_i,\theta_0)$;
these coincide under model conditions but not in general.

To analyse the implications of such a model-robust viewpoint
for our methods, consider first the $\psi_n(t,\theta_0)$
process of Section 2.1. It is clear that its limit $Z_0$
has covariance structure $\min(t_1,t_2)K$.
The Taylor expansion and other arguments of Section 2.2
now show, mutatis mutandem, that $\psi_n(t,\hatt\theta)$
is well approximated with
$\psi_n(t,\theta_0)-tJ_{[nt]}J^{-1}\psi_n(1,\theta_0)$,
with limit $Z(t)=Z_0(t)-tZ_0(1)$. This has covariance
structure $t_1(1-t_2)K$ for $t_1\le t_2$. It follows that
the natural model-robust monitoring process is
$M_n^*(t)=\hatt K^{-1/2}\psi_n(t,\hatt\theta)$,
that is, just like in (2.3), but for model-robust safety
employing $\hatt K$ to estimate the variance matrix of
the score function, rather than $\hatt J$.

In the normal-model case of Example 1, Section 2.4, this
leads to using
$$M_n^*(t)={1\over \sqrt{n}}\sum_{i\le [nt]}
  \pmatrix{1 &\hatt\kappa_1 \cr \hatt\kappa_1 &2+\hatt\kappa_2 \cr}^{-1/2}
  \pmatrix{Z_i \cr Z_i^2-1 \cr},
  \quad {\rm with\ }Z_i=(Y_i-\hatt\mu)/\hatt\sigma $$
instead of the simpler one given there.
Here $\hatt\kappa_1$ and $\hatt\kappa_2$ are sample-based
estimates of skewness and kurtosis.
Similarly, in the Poisson model Example 3 of Section 2.4,
the model-robust viewpoint leads to using
$M_n^*(t)=n^{-1/2}\sum_{i\le [nt]}(Y_i-\bar Y)/\hatt\sigma$,
with $\hatt\sigma=\{n^{-1}\sumin(Y_i-\bar Y)^2\}^{1/2}$
replacing the simpler model-based $\bar Y^{1/2}$.

\subsection
{\sl 7.2. An innovation approach.}
The monitoring process $M_n(t)$ has the
attractive property of converging under parameter constancy
to a vector of $p$ independent Brownian bridges. This property
is lost by stochastic integration, and the limit in distribution of
the weighted monitoring process $\int K_n(s)\,\d M_n(s)$ becomes
less tractable. In \textref{12}{Koning (1999)} and
\textref{8}{Hjort and Koning (2000b)}
this problem is overcome by transforming the monitoring process
$M_n(t)$ into a process $\tilda{M}_n(t)$ which converges
under parameter constancy to a vector of $p$ independent Brownian motions.
This involves innovation transforms in the spirit of
\textref{11}{Khmaladze (1981)}.
Hence, under suitable conditions on $K_n$
the process $\int K_n(s)\,\d\tilda{M}_n(s)$ also converges
to a vector of $p$ independent Brownian motions
(albeit time-transformed). This solution takes a slight toll,
in the sense that the rate of convergence of $\tilda{M}_n(t)$
is a factor $\log n$ less than the rate of convergence of $M_n(t)$.

\subsection
{\sl 7.3. Extensions to Markov and time series models.}
Our methods are not limited to the context of independence
considered in earlier sections of this paper, but have a far
larger generality. In principle, they could be applied
to any statistical model in which we can define
a cumulative score process $\psi_n(t,\theta_0)$
which satisfies (2.2), the starting point for our methodology.
For this purpose martingale central limit theorems
would often suffice for verification.
Presently we verify this property for Markov models,
and then comment shortly on Gau\ss ian autoregressive
type time series models.

For a one-step memory Markov model, taking for technical
expediency the viewpoint that the value $y_1$ of the first
variable $Y_1$ simply is fixed and given, and not
informative for the Markov parameters, we may define
the log-likelihood `at time $t$', that is, corresponding
to the sample $Y_1,\ldots,Y_{[nt]}$, as
$\sum_{i=2}^{[nt]} \log f(Y_{i-1},Y_{i};\theta)$.
Here $f(y,y';\theta)$ is the density of the transition measure
(cf.~\textref{1}{Billingsley (1961)}, p.~4). Defining $u(y,y';\theta)$ as the
first derivative of $f(y,y';\theta)$ with respect to $\theta$,
it can be shown that the cumulative score process
$$\psi_n(t;\theta_0) = {1\over \sqrt{n}} \sum_{i\leq[nt]}
   u(Y_{i-1},Y_i;\theta_0) $$
is a martingale with asymptotic variance function $tJ$,
where $J$ is the information matrix
(see \textref{1}{Billingsley (1961)}, p.~6). Under mild regularity conditions,
application of Rootz\'en's theorem now yields (2.2).

% To my disappointment, the following was all I could
% find in the traditional time series books.
For stationary Gau\ss ian AR, ARMA and ARIMA time series models
we may use the so-called conditional likelihood,
see e.g.~equation (7.1.2) in~\textref{3}{Box, Jenkins and Reinsel (1994)},
to define a cumulative score process which becomes a martingale.
This again yields result (2.2) if Rootz\'en's theorem applies.

\bigskip
{\bf Acknowledgements.}
This work grew out of material originally presented
by one of us at the Oslo May 1998 conference on
discontinuous phenomena in statistics, sponsored by the
the European Science Foundation via their
Highly Structured Stochastic Systems programme.
A.J.K.~is also grateful for hospitality and partial support
in connection with visits to the Department of Mathematics
at the University of Oslo.

\bigskip
\centerline{\bf References}
\def\ref#1#2{{\noindent {[#1]} \hangindent=20pt #2\smallskip}}
% my usual preference
\parindent0pt
\baselineskip11pt
\parskip3pt

\medskip
\ref{1}{%
Billingsley, P. (1961).
{\sl Statistical Inference for Markov Processes.}
University of Chicago press, Chicago.}

\ref{2}{%
Billingsley, P. (1968).
{\sl Convergence of Probability Measures.}
Wiley, New York.}

\ref{3}{%
Box, G.E.P., Jenkins, G.M.~and Reinsel, G.C. (1994).
{\sl Time Series Analysis: Forecasting and Control}
(3rd ed.). Prentice Hall, Englewood Cliffs, N.J.}

%\ref{}{%
%Cox, D.R. (1958).
%{\sl Planning of experiments.}
%John Wiley and Sons, New York.}

\ref{4}{%
Durbin, J. (1973).
Weak convergence of the sample distribution function
when parameters are estimated.
{\sl Annals of Statistics} {\bf 1}, 279--290.}

\ref{5}{%
G\"anssler, P.~and H\"ausler, E. (1979).
Remarks on the functional central limit theorem for martingales.
{\sl Zeitschrift f\"ur Wahrscheinlichkeitstheorie und
verwandte Gebiete} {\bf 50}, 237--243.}

\ref{6}{%
Hjort, N.L. (1990).
Goodness of fit tests in models for life history data
based on cumulative hazard rates.
{\sl Annals of Statistics} {\bf 18}, 1221--1258.}

\ref{7}{%
Hjort, N.L.~and Koning, A.J. (2000a).
Has the density changed?
Unpublished manuscript, Department of Mathematics,
University of Oslo.}

\ref{8}{%
Hjort, N.L.~and Koning, A.J. (2000b).
A general innovation approach towards goodness of fit testing.
Manuscript, in progress.}

\ref{9}{% NOT MENTIONED IN TEXT!
Horv\'ath, L~and Parzen, E. (1994).
Limit theorems for Fisher-score change processes.
In: {\sl Change-Point Problems}. IMS Lecture Notes - Monograph
Series, Vol. 23, 157--169.}

\ref{10}{%
Jacod, J.~and Shiryaev, A.N. (1987).
{\sl Limit Theorems for Stochastic Processes}.
Springer-Verlag, Berlin.}

\ref{11}{%
Khmaladze, E.V. (1981). Martingale approach in the theory of
goodness of fit tests.
{\sl Theory of Probability and its Applications} {\bf 26}, 246--265.}

\ref{12}{%
Koning, A.J. (1999).
Goodness of fit for the constancy of a classical statistical model
over time. Econometric Institute Report 99XX/A, Rotterdam.}
% {\sl Biometrika}, to appear.}

\ref{13}{% SHOULD YEAR BE CHANGED?
Koning, A.J.~and Does, R.J.M.M. (1997).
Cusum charts for preliminary analysis of individual observations.
{\sl Journal of Quality Technology} (to appear).}
% {\sl Econometric Institute Report} {\bf 9727/A}, Rotterdam.}

%\ref{}{%
%Myers, R.H.~and Montgomery, D.C. (1995).
%{\sl Response surface methodology. Process and product optimization
%using designed experiments.}
%John Wiley, New York.}

%\ref{}{%
%Narula, S.C.~and Wellington, J.W. (1977).
%Prediction, linear regression and minimum sum of relative errors.
%{\sl Technometrics} {\bf 19}, 185--190.}

\ref{14}{%
Miller, R.G.~and Siegmund, D. (1982).
Maximally selected chi-square statistics.
{\sl Biometrics} {\bf 38}, 1011--1016.}

\ref{15}{%
de Nationale Ombudsman (1996).
{\sl Openbaar rapport 96.00922.}
de Nationale Ombudsman, 's-Gravenhage.}

\ref{16}{%
Ploberger, W.~and Kr\"amer, W. (1992).
The CUSUM test with OLS residuals.
{\sl Econometrica} {\bf 60}, 271--285.}

\ref{17}{%
Pollard, D. (1984).
{\sl Convergence of Stochastic Processes.}
Springer, New York.}

\ref{18}{%
Ripley, B.D. (1987).
{\sl Stochastic Simulation.} Wiley, New York.}

\ref{19}{%
Rootz\'en, H. (1980).
On the functional limit theorem for martingales.
{\sl Zeitschrift f\"ur Wahrscheinlichkeitstheorie und
verwandte Gebiete} {\bf 51}, 79--93.}

\ref{20}{%
Schruben, L.W. (1982).
Detecting initialization bias in simulation output.
{\sl Operation Research} {\bf 30}, 569--590.}

\ref{21}{%
Schruben, L.W. (1983).
Confidence interval estimation using standardized time series.
{\sl Operation Research} {\bf 31}, 1090--1108.}

\ref{22}{%
Sullivan, J.H.~and Woodall, W.H. (1996).
A control chart for preliminary analysis of individual observations.
{\sl Journal of Quality Technology} {\bf 28}, 265--278.}

\bigskip
\parindent20pt
\baselineskip14pt

\bigskip
\centerline{\bf Appendix: optimal K choice}

\startit
Here an optimality problem raised in Section 4 is solved.
The result leads to the optimal choice of weight functions
$K_{n,j}(s)$ when basing chi squared tests on increments
of $\int_0^tK_{n,j}(s)\,\d M_{n,j}(s)$; see also Section 3.
Specifically, we maximise the $\lambda_j=\lambda_j(K_j)$
parameter of equation (4.4) over all $K_j$ functions.
We may omit the index $j$ in what follows.
We show that $\lambda(K) \le\int_0^1 H^2\,\d s$,
for any nontrivial weight function $K$;
note that this bound is then achieved with $K$ proportional to $H$.

Introduce the vector function $\ell=(KJ_1,\ldots,KJ_m)^\tr$,
where $J_k(t)$ is indicator for $t$ belonging to $I_k$,
and let $L=\ell-c$, where $c=\int_0^1\ell(s)\d s$.
Note next that
$$\int_0^1 LH\,\d s=\int_0^1 \ell H\,\d s=
  \Bigl(\int_{I_1}KH\,\d s,\ldots,\int_{I_m}KH\,\d s\Bigr)^\tr. $$
Moreover, one shows that $\int_0^1 LL^\tr\,\d s$ equals $D-cc^\tr$
(in the notation of Section 3, but with subscript $j$ omitted).
It follows that
$$\lambda(K)=\Bigl(\int_0^1 LH\,\d s\Bigr)^\tr
  \Bigl(\int_0^1 LL^\tr\d s\Bigr)^{-1} \int_0^1 LH\,\d s. $$
That $\lambda(K)\le\int_0^1 H^2\,\d s$ now follows from
a generalised version of the Cauchy--Schwartz inequality:
the matrix
$$\Omega=\pmatrix{\int_0^1 H^2\,\d s &\int_0^1 L^\tr H\,\d s \cr
           \int_0^1 LH\,\d s &\int_0^1 LL^\tr\,\d s \cr}$$
is nonnegative definite, implying that
$\Omega_{11}-\Omega_{12}\Omega_{22}^{-1}\Omega_{21}$ is nonnegative
definite too. This proves the claim.
% \eject

\section
\centerline{\bf Table I: Dutch TBS data}

\smallskip
\def\h{\phantom{1}}
{\smallskip\settabs 12 \columns \baselineskip11pt
\centerline{\sl Number of TBS-sentences}
\smallskip
\+&          &'84 &  '85 &  '86 &  '87 &  '88 &  '89 &  '90 &  '91 & '92 \cr
\smallskip
\+&{\sl Jan} &{\h}1&{\h}7&{\h}8&{\h}7&{\h}8&{\h}9&{\h}8&{\h}5&{\h}4 \cr
\+&{\sl Feb} &{\h}5&  11 &{\h}7&{\h}2&{\h}9&{\h}9&  12 &{\h}6& 12   \cr
\+&{\sl Mar} & 10  &  10 &  14 &{\h}3&  11 &{\h}9&  10 &{\h}8&{\h}3 \cr
\+&{\sl Apr} & 13  &{\h}8&{\h}4&{\h}7&{\h}5&{\h}2&{\h}9&{\h}6& 13   \cr
\+&{\sl May} &{\h}6&{\h}4&{\h}4&{\h}7&{\h}7&{\h}9&  11 &  14 &{\h}6 \cr
\+&{\sl Jun} &{\h}5&{\h}5&{\h}7&{\h}5&{\h}9&{\h}7&{\h}9&{\h}9&{\h}7 \cr
\+&{\sl Jul} & 15  &{\h}6&{\h}8&  10 &{\h}9&  10 &{\h}8&{\h}9& 14   \cr
\+&{\sl Aug} &{\h}5&{\h}8&{\h}2&{\h}4&{\h}3&  11 &{\h}3&{\h}6& 11   \cr
\+&{\sl Sep} &{\h}5&{\h}8&{\h}9&{\h}8&{\h}4&{\h}6&{\h}9&  11 &{\h}8 \cr
\+&{\sl Oct} &{\h}9&{\h}9&{\h}7&{\h}7&{\h}8&{\h}6&{\h}3&  17 &{\h}8 \cr
\+&{\sl Nov} &{\h}6&  16 &  14 &{\h}6&{\h}8&  10 &{\h}7&  14 & 14   \cr
\+&{\sl Dec} & 10  &  14 &  10 &  10 &{\h}9&{\h}6&{\h}6&  12 & 17   \cr

\medskip
\centerline{\sl Number of ended TBS-treatments}
\smallskip
\+&         &'84 & '85 & '86 & '87 & '88 & '89 & '90 & '91 & '92 \cr
\smallskip
\+&{\sl Jan} & 10  &{\h}6&{\h}5&{\h}6& 10  & 10  &{\h}2&{\h}4&{\h}4 \cr
\+&{\sl Feb} &{\h}7&{\h}9&{\h}9& 10  &{\h}7&{\h}8&{\h}2&{\h}4&{\h}6 \cr
\+&{\sl Mar} &{\h}4&{\h}6&{\h}7& 10  &{\h}5& 10  &{\h}6&{\h}9&{\h}6 \cr
\+&{\sl Apr} &{\h}5& 11  &{\h}4&{\h}9&{\h}6&{\h}5&{\h}5&{\h}8&{\h}6 \cr
\+&{\sl May} & 11  &{\h}7&{\h}8&{\h}3& 10  &{\h}8& 12  &{\h}8&{\h}6 \cr
\+&{\sl Jun} &{\h}3&{\h}3&{\h}8&{\h}5&{\h}4&{\h}7&{\h}8&{\h}6&{\h}6 \cr
\+&{\sl Jul} &{\h}8& 11  &{\h}4&{\h}8&{\h}4&{\h}7&{\h}0& 12  &{\h}4 \cr
\+&{\sl Aug} &{\h}6&{\h}5&{\h}5&{\h}7&{\h}3&{\h}6&{\h}4&{\h}4&{\h}7 \cr
\+&{\sl Sep} &{\h}4&{\h}3&{\h}3&{\h}5&{\h}6&{\h}6&{\h}2& 13  &{\h}6 \cr
\+&{\sl Oct} &{\h}2&{\h}7&{\h}4& 10  &{\h}4& 13  &{\h}9& 10  &{\h}2 \cr
\+&{\sl Nov} &{\h}6&{\h}5&{\h}5&{\h}5&{\h}6&{\h}6&{\h}8&{\h}6&{\h}5 \cr
\+&{\sl Dec} & 10  &{\h}8&{\h}8&{\h}6& 12  &{\h}9&{\h}5&{\h}7&{\h}6 \cr
\smallskip}

{\smallskip\narrower\noindent
{\csc Table:} Data given in \textref{15}{de Nationale Ombudsman (1996)},
p.~82--83. See the discussion in Section 6. \smallskip}

\vfill\eject

\baselineskip12pt

\smallskip
\vskip-0.5cm 

\includegraphics[scale=0.55,angle=0]{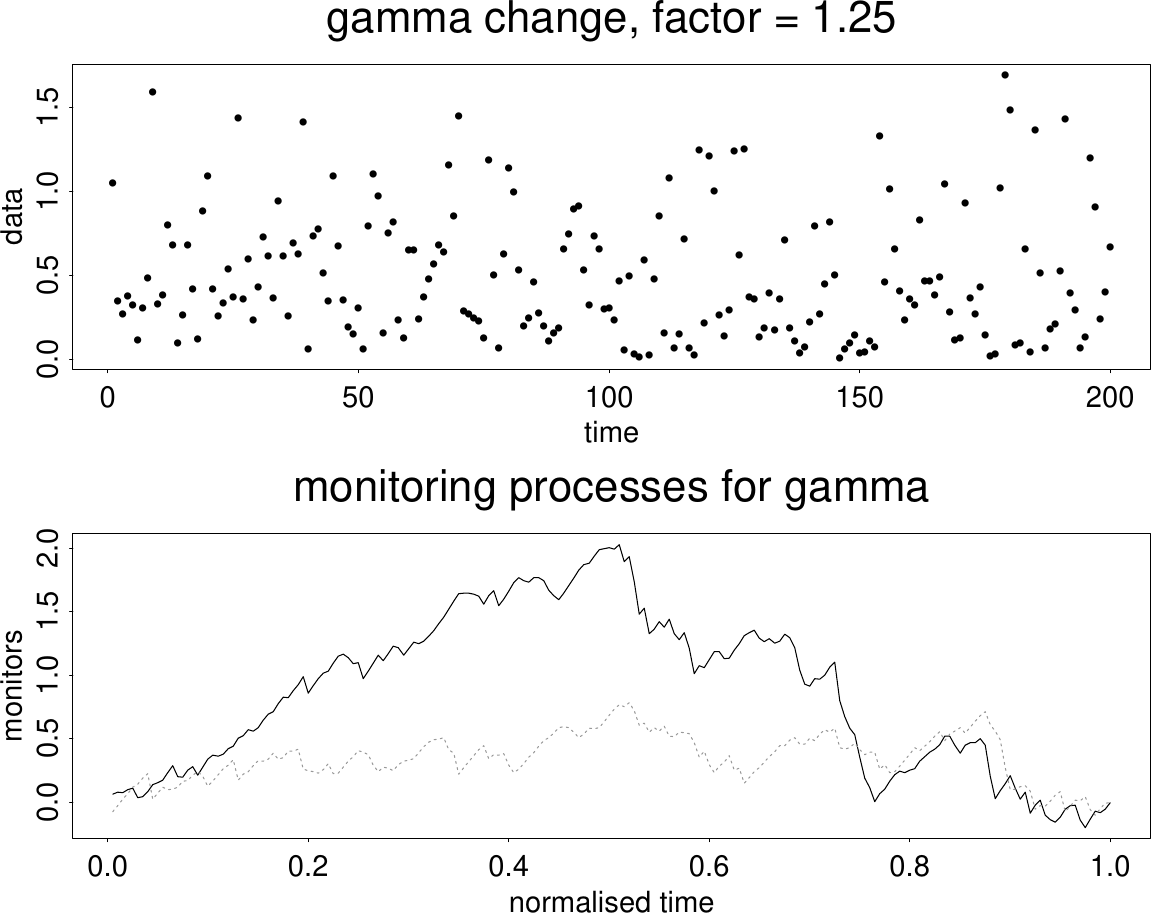}
% \centerline{\includegraphics[height=4.4in,width=3.3in,angle=0]{nils_alex_1-crop.pdf}} 

{\bigskip\narrower\noindent\sl 
{\csc Figure 1.} The first 100 and the following 100 data points 
have been drawn from two different Gamma densities; 
these have the same mean level, but the second has standard 
deviation 1.25 times bigger than that of the first. 
This aspect is barely visible from the data figure,
but is being brought out by the monitoring processes;
the maximum absolute value of the first of these
exceeds the null-distribution 0.95 point of 1.358, for example.
The triangular shape correctly indicates that the non-constancy 
is in form of a break point about half-way through the data.
\smallskip}

\vfill\eject

\smallskip
\vskip-0.5cm 
\centerline{\includegraphics[scale=0.55,angle=0]{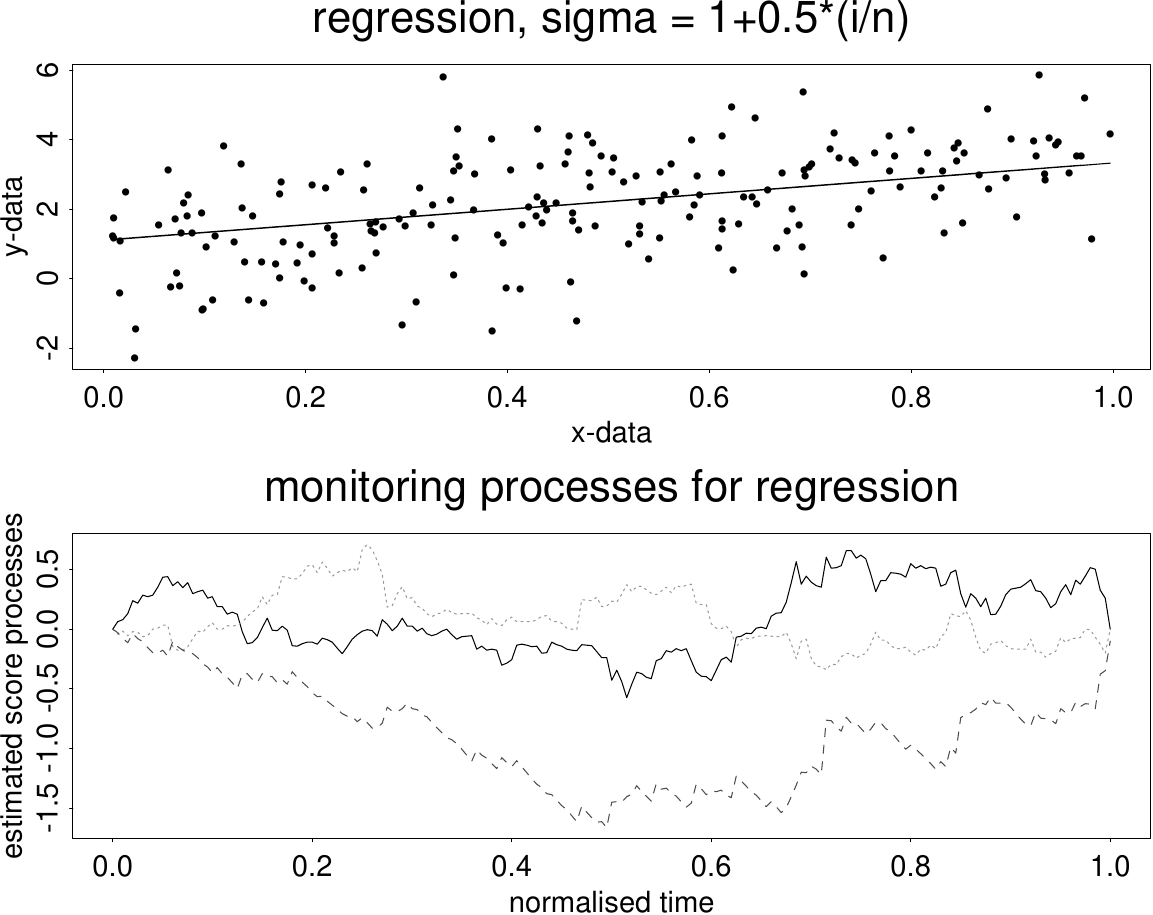}}
% \centerline{\includegraphics[height=4.4in,width=3.3in,angle=0]{nils_alex_2c-crop.pdf}}

{\bigskip\narrower\noindent\sl
{\csc Figure 2.}   
Here $n=200$ pairs are generated by letting 
$x_i$s be independent (and not sorted) uniformly on $(0,1)$
and then using the the $y_i=a+bx_i+\varepsilon_i$ model, 
with normal errors $N(0,\sigma_i^2)$ and using $\sigma_i=1+0.5\,i/n$. 
The monitoring process plots pick up the aspect that 
$\sigma$ is not constant, in that its maximum absolute value 
exceeds the 1.358 value, for example. Also its approximately 
parabolic shape helps identify the type of non-constancy 
of the $\sigma$ parameters.
\smallskip}

\vfill\eject

\smallskip
\vskip-0.5cm 
\centerline{\includegraphics[scale=0.55,angle=0]{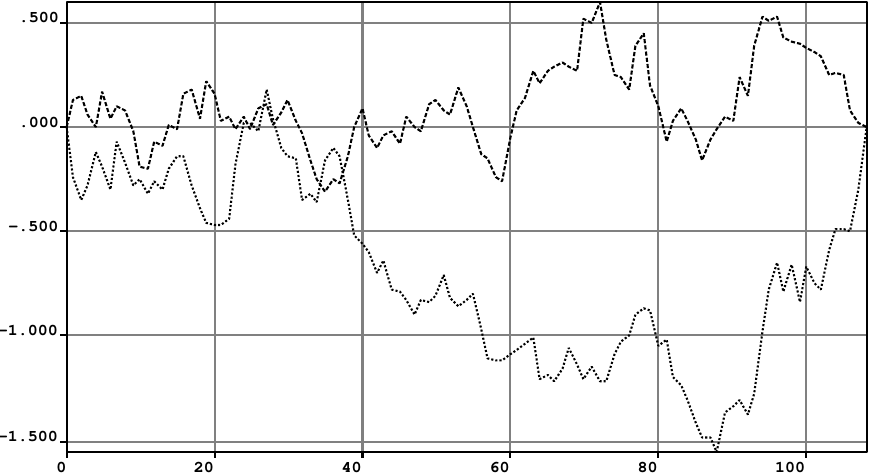}}
%% \centerline{\includegraphics[height=4.4in,width=3.3in,angle=0]{nils_alex_3-crop.pdf}}

{\bigskip\narrower\narrower\noindent\sl 
{\csc Figure 3.} 
Monitoring plots for checking the constancy 
of Poisson parameters, for the two sets of Dutch Ombudsman data
(see Appendix II). The plot for the expected number of TBS-sentences
does not indicate any departure from the hypothesis of constancy,
whereas the plot for the expected number of ended treatments 
indicates that this parameter has not been constant
over the time period studied. The triangular shape indicates 
a sudden decrease around March 1990.
\smallskip}
  
\bye